\input tables
% determine hypertex mode
\newif\iflanl
\openin 1 lanlmac
\ifeof 1 \lanlfalse \else \lanltrue \fi
\closein 1
\iflanl
    \input lanlmac
\else
    \message{[lanlmac not found - use harvmac instead}
    \input harvmac
    \fi
\newif\ifhypertex
\ifx\hyperdef\UnDeFiNeD
    \hypertexfalse
    \message{[HYPERTEX MODE OFF}
    \def\hyperref#1#2#3#4{#4}
    \def\hyperdef#1#2#3#4{#4}
    \def\hypernoname{}
    \def\e@tf@ur#1{}
    \def\eprt#1{{\tt #1}}
    \def\CERN{\centerline{$^b\,$CERN, CH--1211 Geneva 23, Switzerland}}
    \def\wl{W.\ Lerche}
\else
    \hypertextrue
    \message{[HYPERTEX MODE ON}
%hypertex links to xxx.lanl.gov:
%  \def\hth/#1#2#3#4#5#6#7{\special{html:<a
%   href="http://xxx.lanl.gov/abs/hep-th/#1#2#3#4#5#6#7">}
%  {\tt hep-th/#1#2#3#4#5#6#7}\special{html:</a>}}
\def\eprt#1{{\tt
#1}}
\def\CERN{\centerline{

$^b\,$Theory Division, CERN, Geneva, Switzerland}}
\def\wl{
 W.\ Lerche}
\fi
%%%%%%%%%%%%%%%%%%%%%%% %%%%%%%%%%%%%%%%%%%%%%%
\newif\ifdraft

\noblackbox
\catcode`\@=11
\newif\iffrontpage
%%%%%%%%%%%%%%%%%%% %%%%%%%%%%%%%%%%%%%%%%%%%%%%%%%%%%%%%%%%%%%%%%
%%%%% sizes, offsets etc
%%%%%%%%%%%%%%%%%%% %%%%%%%%%%%%%%%%%%%%%%%%%%%%%%%%%%%%%%%%%%%%%%
\ifx\answ\bigans
\def\titleft{\titla}
\magnification=1200\baselineskip=14pt plus 2pt minus 1pt
%
%%%%% unreduced mode: %%%%
%\voffset=0.35truein\hoffset=0.250truein
\advance\hoffset by-0.075truein
\advance\voffset by1.truecm
\hsize=6.15truein\vsize=600.truept\hsbody=\hsize\hstitle=\hsize
\else\let\lr=L
\def\titleft{\titla}
\magnification=1000\baselineskip=14pt plus 2pt minus 1pt
%
%%%%% reduced mode: %%%%%%%
\hoffset=-0.75truein\voffset=-.0truein
%?\hoffset=-.25truein\voffset=-.0truein
\vsize=6.5truein
\hstitle=8.truein\hsbody=4.75truein
\fullhsize=10truein\hsize=\hsbody
\fi
\parskip=4pt plus 15pt minus 1pt
%
%%%%%%%%%%%%%%%%%%% %%%%%%%%%%%%%%%%%%%%%%%%%%%%%%%%%%%%%%%%%%%%%%
%%%%% figures
%%%%%%%%%%%%%%%%%%% %%%%%%%%%%%%%%%%%%%%%%%%%%%%%%%%%%%%%%%%%%%%%%
\newif\iffigureexists
\newif\ifepsfloaded
\def\epsfcheck{
\ifdraft% to speed up
\input epsf\epsfloadedtrue
\else
  \openin 1 epsf
  \ifeof 1 \epsfloadedfalse \else \epsfloadedtrue \fi
  \closein 1
  \ifepsfloaded
    \input epsf
  \else
\immediate\write20{NO EPSF FILE --- FIGURES WILL BE IGNORED}
  \fi
\fi
\def\epsfcheck{}}
\def\checkex#1{
\ifdraft
\figureexistsfalse\immediate%
\write20{Draftmode: figure #1 not included}
\figureexiststrue
\else\relax
    \ifepsfloaded \openin 1 #1
        \ifeof 1
           \figureexistsfalse
  \immediate\write20{FIGURE FILE #1 NOT FOUND}
        \else \figureexiststrue
        \fi \closein 1
    \else \figureexistsfalse
    \fi
\fi}
\def\missbox#1#2{$\vcenter{\hrule
\hbox{\vrule height#1\kern1.truein
\raise.5truein\hbox{#2} \kern1.truein \vrule} \hrule}$}
\def\lfig#1{%  this is to call the figure in the text
\let\labelflag=#1%
\def\numb@rone{#1}%
\ifx\labelflag\UnDeFiNeD%
{\xdef#1{\the\figno}%
\writedef{#1\leftbracket{\the\figno}}%
\global\advance\figno by1%
}\fi{\hyperref{}{figure}{{\numb@rone}}{Fig.{\numb@rone}}}}
\def\figinsert#1#2#3#4{%  this inserts the figure
\epsfcheck\checkex{#4}%
\def\figsize{#3}%
\let\flag=#1\ifx\flag\UnDeFiNeD
{\xdef#1{\the\figno}%
\writedef{#1\leftbracket{\the\figno}}%
\global\advance\figno by1%
}\fi
\goodbreak\midinsert%
\iffigureexists
\centerline{\epsfysize\figsize\epsfbox{#4}}%
\else%
\vskip.05truein
  \ifepsfloaded
  \ifdraft
  \centerline{\missbox\figsize{Draftmode: #4 not included}}%
  \else
  \centerline{\missbox\figsize{#4 not found}}
  \fi
  \else
  \centerline{\missbox\figsize{epsf.tex not found}}
  \fi
\vskip.05truein
\fi%
{\smallskip%
\leftskip 4pc \rightskip 4pc%
\noindent\ninepoint\sl \baselineskip=11pt%
{\bf{\hyperdef\hypernoname{figure}{{#1}}{Fig.{#1}}}:~}#2%
\smallskip}\bigskip\endinsert%
}

\def\boxit#1{\vbox{\hrule\hbox{\vrule\kern8pt
\vbox{\hbox{\kern8pt}\hbox{\vbox{#1}}\hbox{\kern8pt}}
\kern8pt\vrule}\hrule}}
\def\mathboxit#1{\vbox{\hrule\hbox{\vrule\kern8pt\vbox{\kern8pt
\hbox{$\displaystyle #1$}\kern8pt}\kern8pt\vrule}\hrule}}
%
%%%%%%%%%%%%%%%%%%% %%%%%%%%%%%%%%%%%%%%%%%%%%%%%%%%%%%%%%%%%%%%%%
%%%%%  fonts
%%%%%%%%%%%%%%%%%%% %%%%%%%%%%%%%%%%%%%%%%%%%%%%%%%%%%%%%%%%%%%%%%
\font\bigit=cmti10 scaled \magstep1

\font\titla=cmr10 scaled\magstep3
\font\tenmss=cmss10
\font\absmss=cmss10 scaled\magstep1

\newfam\mssfam
\font\footrm=cmr8  \font\footrms=cmr5
\font\footrmss=cmr5   \font\footi=cmmi8
\font\footis=cmmi5   \font\footiss=cmmi5
\font\footsy=cmsy8   \font\footsys=cmsy5
\font\footsyss=cmsy5   \font\footbf=cmbx8
\font\footmss=cmss8
\def\footfont{\def\rm{\fam0\footrm}
\textfont0=\footrm \scriptfont0=\footrms
\scriptscriptfont0=\footrmss
\textfont1=\footi \scriptfont1=\footis
\scriptscriptfont1=\footiss
\textfont2=\footsy \scriptfont2=\footsys
\scriptscriptfont2=\footsyss
\textfont\itfam=\footi \def\it{\fam\itfam\footi}
\textfont\mssfam=\footmss \def\mss{\fam\mssfam\footmss}
\textfont\bffam=\footbf \def\bf{\fam\bffam\footbf} \rm}
\def\tenpoint{\def\rm{\fam0\tenrm}
\textfont0=\tenrm \scriptfont0=\sevenrm
\scriptscriptfont0=\fiverm
\textfont1=\teni  \scriptfont1=\seveni
\scriptscriptfont1=\fivei
\textfont2=\tensy \scriptfont2=\sevensy
\scriptscriptfont2=\fivesy
\textfont\itfam=\tenit \def\it{\fam\itfam\tenit}
\textfont\mssfam=\tenmss \def\mss{\fam\mssfam\tenmss}
\textfont\bffam=\tenbf \def\bf{\fam\bffam\tenbf} \rm}
\ifx\answ\bigans\def\abstractfont{\tenpoint}\else
\def\abstractfont{\def\rm{\fam0\absrm}
\textfont0=\absrm \scriptfont0=\absrms
\scriptscriptfont0=\absrmss
\textfont1=\absi \scriptfont1=\absis
\scriptscriptfont1=\absiss
\textfont2=\abssy \scriptfont2=\abssys
\scriptscriptfont2=\abssyss
\textfont\itfam=\bigit \def\it{\fam\itfam\bigit}
\textfont\mssfam=\absmss \def\mss{\fam\mssfam\absmss}
\textfont\bffam=\absbf \def\bf{\fam\bffam\absbf}\rm}\fi
%
%%%%%%%%%%%%%%%%%%%%%%%%%%%%% %%%%%%%%%%%%%%%%%%%%%%%%%%%%%%%%%
%%%%% footnotes   (adapted from PHYZZX, no hypertext yet)
%%%%%%%%%%%%%%%%%%%%%%%%%%%%% %%%%%%%%%%%%%%%%%%%%%%%%%%%%%%%%%
\def\f@@t{\baselineskip10pt\lineskip0pt\lineskiplimit0pt
\bgroup\aftergroup\@foot\let\next}
\setbox\strutbox=\hbox{\vrule height 8.pt depth 3.5pt width\z@}
\def\vfootnote#1{\insert\footins\bgroup
\baselineskip10pt\footfont
\interlinepenalty=\interfootnotelinepenalty
\floatingpenalty=20000
\splittopskip=\ht\strutbox \boxmaxdepth=\dp\strutbox
\leftskip=24pt \rightskip=\z@skip
\parindent=12pt \parfillskip=0pt plus 1fil
\spaceskip=\z@skip \xspaceskip=\z@skip
\Textindent{$#1$}\footstrut\futurelet\next\fo@t}
\def\Textindent#1{\noindent\llap{#1\enspace}\ignorespaces}
\def\foot{\global\advance\ftno by1%
\attach{\hyperref{}{footnote}{\the\ftno}{\footsymbolgen}}%
\vfootnote{\hyperdef\hypernoname{footnote}{\the\ftno}{\footsymbol}}}%
%   this is for custom footnote marks:
\def\footnote#1{\global\advance\ftno by1%
\attach{\hyperref{}{footnote}{\the\ftno}{#1}}%
\vfootnote{\hyperdef\hypernoname{footnote}{\the\ftno}{#1}}}%
\newcount\lastf@@t           \lastf@@t=-1
\newcount\footsymbolcount    \footsymbolcount=0
\global\newcount\ftno \global\ftno=0
\def\footsymbolgen{\relax\footsym
\global\lastf@@t=\pageno\footsymbol}
\def\footsym{\ifnum\footsymbolcount<0
\global\footsymbolcount=0\fi
{\iffrontpage \else \advance\lastf@@t by 1 \fi
\ifnum\lastf@@t<\pageno \global\footsymbolcount=0
\else \global\advance\footsymbolcount by 1 \fi }
\ifcase\footsymbolcount
\fd@f\dagger\or \fd@f\diamond\or \fd@f\ddagger\or
\fd@f\natural\or \fd@f\ast\or \fd@f\bullet\or
\fd@f\star\or \fd@f\nabla\else \fd@f\dagger
\global\footsymbolcount=0 \fi }
\def\fd@f#1{\xdef\footsymbol{#1}}
\def\space@ver#1{\let\@sf=\empty \ifmmode #1\else \ifhmode
\edef\@sf{\spacefactor=\the\spacefactor}
\unskip${}#1$\relax\fi\fi}
\def\attach#1{\space@ver{\strut^{\mkern 2mu #1}}\@sf}
%
%%%%%%%%%%%%%%%%%%% %%%%%%%%%%%%%%%%%%%%%%%%%%%%%%%%%%%%%%%%%%%%%%
%%%%% References
%%%%%%%%%%%%%%%%%%% %%%%%%%%%%%%%%%%%%%%%%%%%%%%%%%%%%%%%%%%%%%%%%
\newif\ifnref
\def\rrr#1#2{\relax\ifnref\nref#1{#2}\else\ref#1{#2}\fi}
\def\ldf#1#2{\begingroup\obeylines
\gdef#1{\rrr{#1}{#2}}\endgroup\unskip}
\def\nrf#1{\nreftrue{#1}\nreffalse}
\def\doubref#1#2{\refs{{#1},{#2}}}

\nreffalse
\def\refout{\listrefs}

\def\lref{\ldf}

%%%%%%%%%%%%%%%%%%% %%%%%%%%%%%%%%%%%%%%%%%%%%%%%%%%%%%%%%%%%%%%%%
%%%%%%% eq numbering
%%%%%%%%%%%%%%%%%%% %%%%%%%%%%%%%%%%%%%%%%%%%%%%%%%%%%%%%%%%%%%%%%
\def\eqn#1{\xdef #1{(\noexpand\hyperref{}%
{equation}{\secsym\the\meqno}%
{\secsym\the\meqno})}\eqno(\hyperdef\hypernoname{equation}%
{\secsym\the\meqno}{\secsym\the\meqno})\eqlabeL#1%
\writedef{#1\leftbracket#1}\global\advance\meqno by1}
\def\eqnalign#1{\xdef #1{\noexpand\hyperref{}{equation}%
{\secsym\the\meqno}{(\secsym\the\meqno)}}%
\writedef{#1\leftbracket#1}%
\hyperdef\hypernoname{equation}%
{\secsym\the\meqno}{\e@tf@ur#1}\eqlabeL{#1}%
\global\advance\meqno by1}
%old:
\def\eqnalign#1{\xdef #1{(\secsym\the\meqno)}
\writedef{#1\leftbracket#1}%
\global\advance\meqno by1 #1\eqlabeL{#1}}
%
%%%%%%%%%%%%%%%%%%% %%%%%%%%%%%%%%%%%%%%%%%%%%%%%%%%%%%%%%%%%%%%%%
%%%%%%  macros for titlepage, marginnotes, etc
%%%%%%%%%%%%%%%%%%% %%%%%%%%%%%%%%%%%%%%%%%%%%%%%%%%%%%%%%%%%%%%%%

%
\def\chap#1{\newsec{#1}}
\def\chapter#1{\chap{#1}}
\def\sect#1{\subsec{#1}}
\def\section#1{\sect{#1}}
\def\\{\ifnum\lastpenalty=-10000\relax
\else\hfil\penalty-10000\fi\ignorespaces}
\def\note#1{\leavevmode%
\edef\@@marginsf{\spacefactor=\the\spacefactor\relax}%
\ifdraft\strut\vadjust{%
\hbox to0pt{\hskip\hsize%
\ifx\answ\bigans\hskip.1in\else\hskip .1in\fi%
\vbox to0pt{\vskip-\dp
%\vskip4pt
\strutbox\sevenbf\baselineskip=8pt plus 1pt minus 1pt%
\ifx\answ\bigans\hsize=.7in\else\hsize=.35in\fi%
\tolerance=5000 \hbadness=5000%
\leftskip=0pt \rightskip=0pt \everypar={}%
\raggedright\parskip=0pt \parindent=0pt%
\vskip-\ht\strutbox\noindent\strut#1\par%
\vss}\hss}}\fi\@@marginsf\kern-.01cm}
\def\titlepage{%
\frontpagetrue\nopagenumbers\abstractfont%
\hsize=\hstitle\rightline{\vbox{\baselineskip=10pt%
{\abstractfont\pubnum}}}\pageno=0}
\frontpagefalse
\def\pubnum{}
\def\pdate{\number\month/\number\yearltd}
\def\makefootline{\iffrontpage\vskip .27truein
\line{\the\footline}
%\vskip -.1truein\line{\pdate\hfil}
\vskip -.1truein\leftline{\vbox{\baselineskip=10pt%
{\abstractfont\pdate}}}
\else\vskip.5cm\line{\hss \tenrm $-$ \folio\ $-$ \hss}\fi}
\def\title#1{\vskip .7truecm\titlestyle{\titleft #1}}
\def\titlestyle#1{\par\begingroup \interlinepenalty=9999
\leftskip=0.02\hsize plus 0.23\hsize minus 0.02\hsize
\rightskip=\leftskip \parfillskip=0pt
\hyphenpenalty=9000 \exhyphenpenalty=9000
\tolerance=9999 \pretolerance=9000
\spaceskip=0.333em \xspaceskip=0.5em
\noindent #1\par\endgroup }
\def\autskip{\ifx\answ\bigans\vskip.5truecm\else\vskip.1cm\fi}
\def\author#1{\vskip .7in \centerline{#1}}

\def\address#1{\ifx\answ\bigans\vskip.2truecm
\else\vskip.1cm\fi{\it \centerline{#1}}}
\def\abstract#1{
\vskip .5in\vfil\centerline
{\bf Abstract}\penalty1000
{{\smallskip\ifx\answ\bigans\leftskip 2pc \rightskip 2pc
\else\leftskip 5pc \rightskip 5pc\fi
\noindent\abstractfont \baselineskip=12pt
{#1} \smallskip}}
\penalty-1000}
\def\endpage{\tenpoint\supereject\global\hsize=\hsbody%
\frontpagefalse\footline={\hss\tenrm\folio\hss}}
\def\ack{\vskip2.cm\centerline{{\bf Acknowledgements}}}
%
%

%
%%%%%%%%%%%%%%%%%%%%%%%%%%%%% %%%%%%%%%%%%%%%%%%%%%%%%%%%%%%%%%
\def\bfone{\relax{\rm 1\kern-.35em 1}}
\def\inbar{\vrule height1.5ex width.4pt depth0pt}
\def\IC{\relax\,\hbox{$\inbar\kern-.3em{\mss C}$}}
\def\ID{\relax{\rm I\kern-.18em D}}
\def\IF{\relax{\rm I\kern-.18em F}}
\def\IH{\relax{\rm I\kern-.18em H}}
\def\II{\relax{\rm I\kern-.17em I}}
\def\IN{\relax{\rm I\kern-.18em N}}
\def\IP{\relax{\rm I\kern-.18em P}}
\def\IQ{\relax\,\hbox{$\inbar\kern-.3em{\rm Q}$}}
\def\IR{\relax{\rm I\kern-.18em R}}
\font\cmss=cmss10 \font\cmsss=cmss10 at 7pt
\def\ZZ{\relax\ifmmode\mathchoice
{\hbox{\cmss Z\kern-.4em Z}}{\hbox{\cmss Z\kern-.4em Z}}
{\lower.9pt\hbox{\cmsss Z\kern-.4em Z}}
{\lower1.2pt\hbox{\cmsss Z\kern-.4em Z}}\else{\cmss Z\kern-.4em
Z}\fi}

 \def\cB{{\cal B}}
 
\def\cE{{\cal E}}
 
\def\cH{{\cal H}}

 \def\cV{{\cal V}}
\def\nup#1({Nucl.\ Phys.\ $\us {B#1}$\ (}
\def\plt#1({Phys.\ Lett.\ $\us  {#1}$\ (}
\def\cmp#1({Comm.\ Math.\ Phys.\ $\us  {#1}$\ (}
\def\prp#1({Phys.\ Rep.\ $\us  {#1}$\ (}
\def\prl#1({Phys.\ Rev.\ Lett.\ $\us  {#1}$\ (}
\def\prv#1({Phys.\ Rev.\ $\us  {#1}$\ (}
\def\mpl#1({Mod.\ Phys.\ Let.\ $\us  {A#1}$\ (}
\def\ijmp#1({Int.\ J.\ Mod.\ Phys.\ $\us{A#1}$\ (}
\def\tit#1|{{\it #1},\ }
%
%%%%%%%%%%%%%%%%%%%%%%%%%%%%%%%% %%%%%%%%%%%%%%%%%%%%%%%%%%%%%%
%%%%% misc %%%%
%%%%%%%%%%%%%%%%%%%%%%%%%%%%%%%% %%%%%%%%%%%%%%%%%%%%%%%%%%%%%%

%

\def\ni{\noindent}
\def\tilde{\widetilde}

\def\us#1{\underline{#1}}

\def\Coe#1.#2.{{#1\over #2}}

\def\coe#1.#2.{\relax{\textstyle {#1 \over #2}}\displaystyle}

\def\shalf{\relax{\textstyle {1 \over 2}}\displaystyle}

\def\to{\rightarrow}
\def\notin{\hbox{{$\in$}\kern-.51em\hbox{/}}}

%%%%%%%%%%%%%%%%%%%%%%%%%%%

%%%%%%%%%%%%%%%%%%%%%%%%%%%
\catcode`\@=12
%%%%%%%%% end macros  %%%%%%% %%%%%%%%%%%%%%%%%%%%%%%%%%%%%%
%%%%%%%%%%%%%%%%%%%%%%%%%%%% %%%%%%%%%%%%%%%%%%%%%%%%%%%%%

%%%%%%%%%%%%%%%%%%%%%%%%%%%% %%%%%%%%%%%%%%%%%%%%%%%%%%%%%
% references: 
%%%%%%%%%%%%%%%%%%%%%%%%%%%% %%%%%%%%%%%%%%%%%%%%%%%%%%%%%

%\def\eprt#1{{\tt #1}}
\def\nihil#1{{\sl #1}}
\def\br{\hfill\break}

\def\ijmp {{Int. J. Mod. Phys.\ }{\bf A}}

\lref\COFKM{P.\ Candelas, X.\ De La Ossa, A.\ Font, 
S.\ Katz and D.R.\ Morrison, 
\nihil{Mirror symmetry for two parameter models.\ I,}
 Nucl.\ Phys.\ {\bf B416} 481 (1994), 
\eprt{hep-th/9308083}. 
%%CITATION = NUPHA,B416,481;%%
}

\lref\DDCR{D.\ Diaconescu and C.\ R\"omelsberger, 
\nihil{D-branes and bundles on elliptic fibrations,}
Nucl.\ Phys.\ {\bf B574} 245 (2000),
\eprt{hep-th/9910172}. 
%%CITATION = HEP-TH 9910172;%%
}

\lref\BDLR{I.\ Brunner, M.R.\ Douglas, A.\ Lawrence and 
C.\ R\"omelsberger, 
\nihil{D-branes on the quintic,}
\eprt{hep-th/9906200}. 
%%CITATION = HEP-TH 9906200;%%
}

\lref\GS{
{M.\ Gutperle and Y.\ Satoh, 
\nihil{D-branes in Gepner models and supersymmetry,}
 Nucl.\ Phys.\ {\bf B543} 73 (1999), 
\eprt{hep-th/9808080};
%%CITATION = NUPHA,B543,73;%%
}
{
\nihil{D0-branes in Gepner models and N = 2 black holes,}
 Nucl.\ Phys.\ {\bf B555} 477 (1999), 
\eprt{hep-th/9902120}. 
%%CITATION = NUPHA,B555,477;%%
}
}

\lref\RS{A.\ Recknagel and V.\ Schomerus, 
\nihil{D-branes in Gepner models,}
 Nucl.\ Phys.\ {\bf B531} 185 (1998), 
\eprt{hep-th/9712186}. 
%%CITATION = NUPHA,B531,185;%%
}

\lref\RSb{
{A.\ Recknagel and V.\ Schomerus, 
\nihil{Boundary deformation theory and moduli spaces of D-branes,}
 Nucl.\ Phys.\ {\bf B545} 233 (1999), 
\eprt{hep-th/9811237};
%%CITATION = NUPHA,B545,233;%%
}
{ 
\nihil{Moduli spaces of D-branes in CFT-backgrounds,}
\eprt{hep-th/9903139}. 
%%CITATION = HEP-TH 9903139;%%
}
}

\lref\MD{M.R.\ Douglas, 
\nihil{Topics in D-geometry,}
Class.\ Quant.\ Grav.\ {\bf 17} 1057 (2000),
\eprt{hep-th/9910170}. 
%%CITATION = HEP-TH 9910170;%%
}

\lref\OOY{H.\ Ooguri, Y.\ Oz and Z.\ Yin, 
\nihil{D-branes on Calabi-Yau spaces and their mirrors,}
 Nucl.\ Phys.\ {\bf B477} 407 (1996), 
\eprt{hep-th/9606112}. 
%%CITATION = HEP-TH 9606112;%%
}

\lref\DDJG{D.\ Diaconescu and J.\ Gomis, 
\nihil{Fractional branes and boundary states in orbifold theories,}
\eprt{hep-th/9906242}. 
%%CITATION = HEP-TH 9906242;%%
}

\lref\GJS{S.\ Govindarajan, T.\ Jayaraman and T.\ Sarkar, 
\nihil{World sheet approaches to D-branes on supersymmetric cycles,}
Nucl.\ Phys.\ {\bf B580} 519 (2000),
\eprt{hep-th/9907131}. 
%%CITATION = HEP-TH 9907131;%%
}

\lref\FS{J.\ Fuchs and C.\ Schweigert, 
\nihil{Branes: From free fields to general backgrounds,}
 Nucl.\ Phys.\ {\bf B530} 99 (1998), 
\eprt{hep-th/9712257}. 
%%CITATION = HEP-TH 9712257;%%
}

\lref\KLLW{P.\ Kaste, W.\ Lerche, C.A.\ L\"utken and J.\ Walcher, 
\nihil{D-branes on K3-fibrations,}
\eprt{hep-th/9912147}. 
%%CITATION = HEP-TH 9912147.%% 
%\href{\wwwspires?eprint=HEP-TH/9912147}{SPIRES}
}

\lref\ES{E.\ Scheidegger, 
\nihil{D-branes on some one- and two-parameter Calabi-Yau hypersurfaces,}
JHEP {\bf 0004} 003 (2000),
\eprt{hep-th/9912188}. 
%%CITATION = HEP-TH 9912188;%% 
%\href{\wwwspires?eprint=HEP-TH/9912188}{SPIRES}
}

\lref\IBVS{I.\ Brunner and V.\ Schomerus, 
\nihil{D-branes at singular curves of Calabi-Yau compactifications,}
 JHEP {\bf 0004} 020 (2000), 
\eprt{hep-th/0001132}. 
%%CITATION = HEP-TH 0001132;%%
}

\lref\DFR{M.R.\ Douglas, B.\ Fiol and C.\ R\"omelsberger, 
\nihil{The spectrum of BPS branes on a noncompact Calabi-Yau,}
\eprt{hep-th/0003263}. 
%%CITATION = HEP-TH 0003263;%%
}

\lref\DFRa{M.R.\ Douglas, B.\ Fiol and C.\ R\"omelsberger, 
\nihil{Stability and BPS branes,}
\eprt{hep-th/0002037}. 
%%CITATION = HEP-TH 0002037;%%
}

\lref\BS{M.\ Bershadsky and V.\ Sadov, 
\nihil{F-theory on K3$\,\times\,$K3 and instantons on 7-branes,}
 Nucl.\ Phys.\ {\bf B510} 232 (1998), 
\eprt{hep-th/9703194}. 
%%CITATION = HEP-TH 9703194
}

\lref\OGRADY{K.\ O'Grady, 
\nihil{Desingularized moduli spaces of sheaves on a K3, I, II},
\eprt{alg-geom/9708009} and \eprt{math.AG/9805099}.
}

\lref\MUKAI{S.\ Mukai,
\nihil{Moduli of vector bundles on $K3$ surfaces and symplectic
manifolds}, Sugaku Expositions Vol.\ 1 (1988) 139.}

\lref\MNMN{M.\ Naka and M.\ Nozaki, 
\nihil{Boundary states in Gepner models,}
 JHEP {\bf 0005} 027 (2000), 
\eprt{hep-th/0001037}. 
%%CITATION = HEP-TH 0001037;%%
}

\lref\KS{K.\ Sugiyama, 
\nihil{Comments on central charge of topological 
sigma model with Calabi-Yau target space,}
\eprt{hep-th/0003166}. 
%%CITATION = HEP-TH 0003166;%%
}

\lref\PSS{
{G.\ Pradisi, A.\ Sagnotti and Y.\ S.\ Stanev, 
\nihil{Completeness conditions for boundary 
operators in 2D conformal field theory,}
 Phys.\ Lett.\ {\bf B381} 97 (1996), 
\eprt{hep-th/9603097};\br 
%%CITATION = HEP-TH 9603097;%%
}
{A.\ Sagnotti and Y.S.\ Stanev, 
\nihil{Open descendants in conformal field theory,}
 Fortsch.\ Phys.\ {\bf 44} 585 (1996), 
\eprt{hep-th/9605042}. 
%%CITATION = HEP-TH 9605042;%%
}}

\lref\VW{
{C.\ Vafa and E.\ Witten, 
\nihil{A Strong coupling test of S duality,}
 Nucl.\ Phys.\ {\bf B431} 3 (1994), 
\eprt{hep-th/9408074}; \br
%%CITATION = HEP-TH 9408074;%%
}
{C.\ Vafa, 
\nihil{Instantons on D-branes,}
 Nucl.\ Phys.\ {\bf B463} 435 (1996), 
\eprt{hep-th/9512078}. 
%%CITATION = HEP-TH 9512078;%%
}}

\lref\GIES{D.\ Gieseker, 
\nihil{On the moduli of vector bundles on an algebraic surface},
Ann.\ Math.\ {\bf 106} 45 (1977).
}  

\lref\FSW{J.\ Fuchs, C.\ Schweigert and J.\ Walcher, 
\nihil{Projections in string theory and 
boundary states for Gepner models,}
\eprt{hep-th/0003298}. 
%%CITATION = HEP-TH 0003298;%%
}

\lref\GRPL{B.R.\ Greene and M.R.\ Plesser, 
\nihil{Duality In Calabi-Yau Moduli Space,}
 Nucl.\ Phys.\ {\bf B338} 15 (1990). 
%%CITATION = NUPHA,B338,15;%%
}

\lref\BSSY{
{A.N.\ Schellekens and S.\ Yankielowicz, 
\nihil{Extended chiral algebras and modular invariant partition functions,}
 Nucl.\ Phys.\ {\bf B327} 673 (1989); \br 
%%CITATION = NUPHA,B327,673;%%
}
For a review, see:
{A.N.\ Schellekens, 
\nihil{Open strings, simple currents and fixed points,}
\eprt{hep-th/0001198}. 
%%CITATION = HEP-TH 0001198;%%
}
}

\lref\HM{J.A.\ Harvey and G.\ Moore, 
\nihil{Algebras, BPS states, and strings,}
 Nucl.\ Phys.\ {\bf B463} 315 (1996), 
\eprt{hep-th/9510182}. 
%%CITATION = HEP-TH 9510182;%%
}

\lref\DM{M.R.\ Douglas and G.\ Moore, 
\nihil{D-branes, quivers, and ALE instantons,}
\eprt{hep-th/9603167}. 
%%CITATION = HEP-TH 9603167;%%
}

\lref\CV{C.\ Vafa, 
\nihil{Instantons on D-branes,}
 Nucl.\ Phys.\ {\bf B463} 435 (1996), 
\eprt{hep-th/9512078}. 
%%CITATION = HEP-TH 9512078;%%
}

\lref\DOFI{
{M.R.\ Douglas, 
\nihil{D-branes and discrete torsion,}
\eprt{hep-th/9807235}; 
%%CITATION = HEP-TH 9807235;%%
}\br
{M.R.\ Douglas and B.\ Fiol,
\nihil{D-branes and discrete torsion II,}
  \eprt{hep-th/9903031}.
%%CITATION = HEP-TH 9903031;%%
}
}

\lref\GOMI{J.\ Gomis,
\nihil{D-branes on orbifolds with discrete torsion 
and topological obstruction,}
   JHEP {\bf 0005} 024 (2000),
  \eprt{hep-th/0001200}.
%%%CITATION = HEP-TH 0001200;%%
}

\lref\FUSSF{J.\ Fuchs, A.N.\ Schellekens, and C.\ Schweigert,
\nihil{A matrix $S$ for all simple current extensions,}
 Nucl.\ Phys.\ {\bf B473} (1996) 323,
\eprt{hep-th/9601078}.
}
           
\lref\GEPNE{D.\ Gepner, 
\nihil{Space-time supersymmetry in compactified string theory and 
superconformal models,}
 Nucl.\ Phys.\ {\bf B296} (1988) 757.
}
          
\lref\FUSCE{J.\ Fuchs and C.\ Schweigert,
\nihil{A classifying algebra for boundary conditions,}
 Phys.\ Lett.\ {\bf B414} (1997) 251,
\eprt{hep-th/9708141}.
}

\lref\HUSSB{L.R.\ Huiszoon, A.N.\ Schellekens, and N.\ Sousa,
\nihil{Open descendants of non-diagonal invariants,}
Nucl.\ Phys.\ {\bf B575} 401 (2000),
 \eprt{hep-th/9911229}.
}

\lref\BANTG{P.\ Bantay,
\nihil{The untwisted stabilizer in simple current extensions,}
 Phys.\ Lett.\ {\bf B396} (1997) 183.
}

\lref\KRSC{M.\ Kreuzer and A.N.\ Schellekens,
\nihil{Simple currents versus orbifolds with discrete torsion 
-- a complete classification,}
 Nucl.\ Phys.\ {\bf B411} (1994) 97.
}

\lref\TOAPP{J.\ Fuchs, L.R.\ Huiszoon, A.N.\ Schellekens, C.\ Schweigert
and J. Walcher, to appear.}

\lref\FW{D.S.\ Freed and E.\ Witten, 
\nihil{Anomalies in string theory with D-branes,}
\eprt{hep-th/9907189}. 
%%CITATION = HEP-TH 9907189;%%
}

\lref\fract{
{D.\ Diaconescu, M.R.\ Douglas and J.\ Gomis, 
\nihil{Fractional branes and wrapped branes,}
 JHEP {\bf 9802} 013 (1998), 
\eprt{hep-th/9712230}; 
%%CITATION = HEP-TH 9712230;%%
}\br
{D.\ Diaconescu and J.\ Gomis, 
\nihil{Fractional branes and boundary states in orbifold theories,}
\eprt{hep-th/9906242}. 
%%CITATION = HEP-TH 9906242;%%
}
}

\lref\WKTH{E.\ Witten, 
\nihil{D-branes and K-theory,}
 JHEP {\bf 9812} 019 (1998), 
\eprt{hep-th/9810188}. 
%%CITATION = HEP-TH 9810188;%%
}

\lref\BBACK{
{A.\ Kapustin, 
\nihil{D-branes in a topologically nontrivial B-field,}
\eprt{hep-th/9909089};
%%CITATION = HEP-TH 9909089;%%
}
{P.\ Bouwknegt and V.\ Mathai, 
\nihil{D-branes, B-fields and twisted K-theory,}
 JHEP {\bf 0003} 007 (2000), 
\eprt{hep-th/0002023}. 
%%CITATION = HEP-TH 0002023;%%
}
}
\lref\EDMD{D.\ Diaconescu and M.R.\ Douglas, 
\nihil{D-branes on stringy Calabi-Yau manifolds,}
\eprt{hep-th/0006224}. 
%%CITATION = HEP-TH 0006224;%%
}

\lref\BFMM{B.\ Fiol and M.\ Marino, 
\nihil{BPS states and algebras from quivers,}
\eprt{hep-th/0006189}. 
%%CITATION = HEP-TH 0006189;%%
}

\lref\VW{C.\ Vafa and E.\ Witten, 
\nihil{On orbifolds with discrete torsion,}
 J.\ Geom.\ Phys.\ {\bf 15} 189 (1995), 
\eprt{hep-th/9409188}. 
%%CITATION = HEP-TH 9409188;%%
}

\lref\CVTORS{C.\ Vafa, 
\nihil{Modular invariance and discrete torsion on orbifolds,}
 Nucl.\ Phys.\ {\bf B273} 592 (1986). 
%%CITATION = NUPHA,B273,592;%%
}

\lref\AMG{P.S.\ Aspinwall, D.R.\ Morrison and M.\ Gross, 
\nihil{Stable singularities in string theory,}
 Commun.\ Math.\ Phys.\ {\bf 178} 115 (1996), 
\eprt{hep-th/9503208}. 
%%CITATION = HEP-TH 9503208;%%
}

%%%%%%% paper  specific macros

\def\frac#1#2{{#1\over #2}}

\def\cals{{{\cal S}}}
\def\calu{{{\cal U}}}

\def\zet{\ZZ}

\def\ee{{\rm e}}
\def\ii{{\rm i}}
\def\text#1{{\rm #1 }}
\def\rr#1{\Phi_{RR}^{(#1)}}
\def\cS{\cals}
\def\cU{\calu}
\def\N{{\cal N}}

\def\GCD{{\rm gcd}}

\def\IG{\relax\,\hbox{$\inbar\kern-.3em{\mss G}$}}

    \def\pk{P.\ Kaste}
    \def\cl{C.A.\ L\"utken}
    \def\jw{J.\ Walcher}
     \def\jf{J.\ Fuchs}
     \def\cs{C.\ Schweigert}

%%%%%%%%%%%%%%%%%%%%%%%%%%%% %%%%%%%%%%%%%%%%%%%%%%%%%%%%%

%\draft

\def\pubnum{
\hbox{CERN-TH/2000-164}
\hbox{PAR-LPTHE 00-030}
\hbox{ETH-TH/00-8}
\hbox{hep-th/0007145}
\hbox{}}
\def\pdate{}
\titlepage
%\vskip2.cm
\title{{\titlefont Boundary Fixed Points, Enhanced Gauge Symmetry
 and Singular Bundles on K3}}
\vskip -1.cm
\autskip
\author{
\jf$\,^a$,
\pk$\,^b$, 
\wl$\,^{b,c}$,
\cl$\,^{b,}$\footnote{1}{On leave from Dept.\ of 
Physics, University of Oslo, N-0316 Oslo, Norway},
\cs$\,^c$
and  
\jw$\,^{b,}$\footnote{2}{Also at Institut f\"ur Theoretische Physik, 
ETH-H\"onggerberg, CH-8093 Z\"urich, Switzerland} 
}  \vskip0.7truecm
\centerline{$^a\,$Institutionen f\"or Fysik, 
Universitetsgatan 1, S-65188 Karlstad, Sweden}
\vskip.2truecm
\CERN
\vskip.2truecm
\centerline{$^c\,$LPTHE, Universit\'e Paris VI, 
4 place Jussieu, F-75252 Paris, Cedex 05, France}
\vskip-.4truecm

\abstract{
We investigate certain fixed points in the boundary conformal field
theory representation of type IIA $D$-branes on Gepner points of $K3$.
They correspond geometrically to degenerate brane configurations, and
physically lead to enhanced gauge symmetries on the world-volume.
Non-abelian gauge groups arise if the stabilizer  group of the fixed
points is realized projectively, which is similar to $D$-branes on
orbifolds with discrete torsion. Moreover, the fixed point boundary
states can be resolved into several  irreducible components.  These
correspond to bound states at threshold and can be viewed as
(non-locally free)  sub-sheaves of semi-stable sheaves. Thus, the BCFT
fixed points appear to carry two-fold geometrical information: on the
one hand they probe the boundary of the instanton moduli space on $K3$,
on the other hand they probe discrete torsion in $D$-geometry.
}

\vfil
%\vskip 1.cm
\ni {CERN-TH/2000-164}\hfill\break
\ni July 2000
\endpage
\baselineskip=14pt plus 2pt minus 1pt

\sequentialequations

%%%%%%%%%%%%%%%%%%%%%%%%%%%%%%%%%%%%%%%%%%%%%
\chapter{Introduction}
%%%%%%%%%%%%%%%%%%%%%%%%%%%%%%%%%%%%%%%%%%%%%

Conformal field theory on world-sheets with boundary (BCFT) has 
proven to be a powerful tool for studying the quantum geometry 
of $D$-branes. In particular, investigating exactly solvable tensor 
products of $\N\,{=}\,2$ minimal models
\doubref\RS\FS\ has provided important
insight in the $D$-geometry of Calabi-Yau threefolds, in the domain of
strong quantum corrections 
\nrf{\BDLR\MD\DDCR\KLLW\ES\IBVS\MNMN\DFR\KS\EDMD} 
\refs{\BDLR-\EDMD}. The purpose of the
present letter is to investigate extra gauge
symmetries stemming from certain fixed points in the BCFT.
As these arise already in type II compactifications on $K3$, we
will focus here on this particularly simple situation
where all properly constructed BCFT states should have a well-defined,
classical geometric interpretation; however, most of our CFT 
considerations directly apply also to general $n$-folds.

The problem can be stated as follows. Using the methods developed in
\BDLR, given some Fermat (``Gepner'') 
point on a $K3$ surface one can easily find a
list of boundary states that correspond to $D$-brane configurations
$\cE$ with $RR$-charges\foot{As usual, $c_i$ denote
Chern classes and rk the rank of the corresponding bundle or sheaf.}
$$
v(\cE)\ =\ (q_4,q_2,q_0)\ \equiv\ ({\rm rk},c_1,r+\shalf
{c_1}^2\!-\!c_2)\,\in H^{{\rm even}}(K3,\ZZ)\ .
\eqn\Mukdef
$$
It is in fact possible to extract more bundle data than just the
charges from the BCFT, like for instance the number of moduli of  a
configuration. For some configurations it was found in
\doubref\DDCR\KLLW\ that the number $\nu$ of vacuum states in the  open
string sector is larger than one. Physically, this corresponds to extra
gauge fields on the world-volume and mathematically to a  degenerate
bundle or sheaf. If there are $\tilde \nu\,{>}\,1$ $U(1)$
factors in the gauge group (where $1\leq\tilde\nu\leq\nu$), such
configurations should be considered as reducible because each $U(1)$
corresponds to an independent center-of-mass degree of freedom of a
multi-brane system. There is no fundamental distinction between a
multi-brane system and a bound state at threshold with the same overall
charges, as these configurations live on the same continuous branch of
moduli space.

Our aim is to investigate such degenerate configurations with 
$\nu\,{>}\,1$, which correspond to singular boundary points of the 
instanton moduli space where extra gauge fields appear. In
particular, we would like to find what the charge vectors $v^{(i)}$
($i\,{=}\,1,...,\tilde\nu$) of the individual components of a reducible
configuration are. This first of all requires understanding the CFT
origin of the non-trivial multiplicities. As we will see, they are
rooted in certain ``simple current'' fixed points that generically
appear on the boundary but not in the bulk CFT. This reflects the fact
that the relevant geometric singularities do not 
appear on the manifold, but rather in bundles over it. 
Upon resolving the fixed points, we will obtain the brane charges $v^{(i)}$ 
of the irreducible components; they turn out to have
a canonical description in terms of sub-sheaves of semi-stable sheaves.

We will also find that while $\nu$ is given by the order of 
the stabilizer $\cS$ of the fixed point, the number $\tilde\nu$
of irreducible components is given by the order of the so-called
\FUSSF\ untwisted stabilizer $\cU$. If the stabilizer is realized 
only projectively on the fixed point, then 
${\nu/\tilde\nu}\,{\equiv}\,N^2\,{>}\,1$, which leads to enhanced 
$U(N)$ gauge symmetry on the world-volume. This is analogous to
$D$-branes on orbifolds with discrete torsion 
\doubref\DOFI\GOMI, and provides an
interesting mechanism for obtaining non-abelian gauge symmetries
in type II string compactifications, within the conformal 
field theory of $\N\!=\!2$ minimal models. 

%%%%%%%%%%%%%%%%%%%%%%%%%%%%%%%%%%%%%%%%%%%%%%%%%%%%%%%%%%%
\chapter{Fixed points in $B$-type $\N\,{=}\,2$ boundary CFT}
%%%%%%%%%%%%%%%%%%%%%%%%%%%%%%%%%%%%%%%%%%%%%%%%%%%%%%%%%%%

We consider the CFT describing the internal part of a Gepner model for
an $n$-fold (where $n$ is the complex dimension of the Calabi-Yau
space)  at the Fermat point. It is constructed out of the tensor
product of $r$  $\N\,{=}\,2$ minimal models with levels $k_i$, suitably
projected to  ensure worldsheet supersymmetry and to allow (after the
GSO projection) for supersymmetry in the external spacetime \GEPNE.
While $A$- and  $B$-type boundary conditions \OOY\ for such models have
been studied over the past few years, the resolution of fixed points
under these projections has not yet been fully worked out. 

More
concretely, for $A$-type states (associated to real submanifolds) the
algebraic problems associated with fixed points were pointed out in
\GJS, analyzed in an example in \IBVS,  and solved in \FSW. For
$B$-type states,  associated to holomorphic geometry, fixed point
phenomena were first noticed in a geometrical context in \DDCR.
Specifically, recall that the $B$-type states in \RS\ were (partially)
labelled by integers $\vec L\,{=}\,(L_1,...,L_r)$, where $0\,{\leq}\,
L_i\,{\leq}\,[k_i/2]$. It was observed in \refs{\DDCR,\KLLW,\IBVS}\
that  whenever a label $L_i$ reaches $k_i/2$, then by virtue of field
identifications there is an extra copy of the vacuum state contributing
to open string amplitudes. Each such vacuum state corresponds to a
gauge field on the brane world-volume \BDLR. The details depend on
whether $n\,{+}\,r$ is  even or odd \KLLW, and altogether it is found
that the total number of such vacuum states is:
$$
\nu\ =\ 2^{\tilde\ell}\ ,\qquad\ \
\tilde\ell\ =\ \cases{
\ell & $n+r$ odd \cr
{\ell-1} & $n+r$ even, $\ell>0$ \cr
0 & $n+r$\ even, $\ell=0$\ , \cr
}\eqn\nucases
$$
where $\ell$ is the number of $L_i$ equal to $k_i/2$. When all $k_j$ are odd,
as for example in the case of the quintic threefold, then $\nu$ is always 
equal to one and the phenomena that we are going to discuss do not appear.

The peculiarity of labels $L_i\,{=}\,k_i/2$ has been recognized before 
in a different context, namely boundary operators in su$(2)$ 
WZW models \PSS\ with $D_{\rm odd}$-type modular invariant.
It has been traced to the fixed point of the simple current that
generates the modular invariant \refs{\PSS,\FUSCE,\HUSSB}.
This simple current has non-integer conformal dimension and 
therefore does not lead to a fixed point in the bulk theory.
However it leads to a fixed point in the boundary CFT, and as a
consequence the boundary state splits into a pair of states. 
Our aim is to resolve the analogous fixed points in $\N\,{=}\,2$ Gepner
models, and find what the charges of the resolved boundary states are.

%%%%%%%%%%%%%%%%%%%%%%%%%%%%%%%%%%%%%%%%%%%%%%%%%%%%%%%%%%%%%%%%%%%
\section{Simple currents in the Gepner-Greene-Plesser construction}
%%%%%%%%%%%%%%%%%%%%%%%%%%%%%%%%%%%%%%%%%%%%%%%%%%%%%%%%%%%%%%%%%%%

At the level of chiral CFT, the problem of constructing B-type boundary
conditions is characterized by both an increase in chiral symmetry
(guaranteeing integrality of $U(1)$ charge) and a partial breaking of
the chiral symmetry  (because of ``twisted gluing conditions'' \RS, or
non-trivial  ``automorphism type'' \FS).   To deal with these
complications in a direct manner would require  the development of new
CFT techniques.   However, mirror symmetry exchanges the $A$- and
$B$-type boundary conditions, so that we can more easily construct
$B$-type states as $A$-type states in the mirror model. As is
well-known, the mirror model can be obtained, according  to the
Greene-Plesser \GRPL\ construction, by modding out all phase
symmetries, and BCFT constructions allowing to deal with such a
situation are available right now.

For the closed string spectrum it is sufficient to  study the action of the
orbifold group on the primary fields of the theory.  However, for obtaining
modular data, and a fortiori boundary conditions  and the open string spectrum,
it is essential to understand the chiral  realization of the
symmetries. This can be done with simple current
techniques. We will not describe those techniques in any detail  here,
but rather refer to ref.\ \BSSY\ for simple currents in general, and
to \FSW\ for the application of simple currents to the construction of
A-type boundary conditions in Gepner models, as well as for further
references. The general theory of boundary conditions in simple current
invariants will appear elsewhere \TOAPP. (See also the appendix of
the present paper.)

For a single $\N\,{=}\,2$ minimal model at level $k$ (whose primary 
fields are labelled, up to field identification, by $(l,m,s)$), 
the most important simple currents are listed in Table~1.
\vskip .3cm
\begintable
simple current | $(l,m,s)$ & order & conf.\ weight  \cr
$v$ | (0,0,2)   & 2            & 3/2      \nr
$s$ | (0,1,1) & \ $4h$ or $2h$ & $k/8h$   \nr
$p$ | (0,2,0)   & $h$          & $k+1/h$   \nr
$f$ | ($k$,0,0) & 2            & $k/4$  
\endtable
%\vskip-10pt
\noindent{\bf Table 1:}
{\sl
The most important simple currents of $\N\,{=}\,2$ minimal models,
for odd or even level $k$, and with $h=k+2$.
} 

In more familiar terms, we note that $v$ is just the primary field that
contains the world-sheet supercurrent(s), $s$ is the primary field that
contains the spectral flow operator, while the monodromies of $p$ 
give the phase symmetries. The simple current $f$ is distinguished
because it is the only one with potential fixed points. It acts on 
primary fields as $f(l,m,s)=(k-l,m,s)$, and thus, if $k$ is even, 
then $f$ fixes all fields of the form $(l,m,s)\,{=}\,(k/2,m,s)$. 
By field identification we can write:
$$
f=(k,0,0) \equiv (0,h,2) =p^{h/2}v\ .
\eqn\fcurrent
$$

When forming the tensor product of minimal models, we add a subscript $i$ to 
indicate the factor. The Calabi-Yau projection, which turns the 
tensor product into the exact solution of the Calabi-Yau
sigma model, can be thought of as a simple current 
extension by the currents $w_i\,{=}\,v_1v_i$, $i\,{=}\,2,...,r$ and by 
$u\,{=}\,v_1^{n+r} \prod_i p_i$. To obtain the mirror model, one must in 
addition include into the simple current group all invariant phase 
symmetries, i.e., all combinations $v_1^\epsilon\prod {p_i}^{\pi_i}$ 
that satisfy
$$
\sum_i \pi_i/h_i + \epsilon/2 \in\ZZ\ ,
\eqn\phases
$$
where $\pi_i\,{=}\,0,...,h_i{-}1$ and $\epsilon=0$ if $n+r$ is even, and
$\epsilon\,{=}\,0,1$ if $n+r$ is odd.\foot{The possibility of having 
$\epsilon\,{\not=}\,0$ is usually neglected in the literature, because it 
is irrelevant for the computation of the closed string NS spectrum. This 
is no longer true in the open string sector. 
}

As a first step in analyzing the fixed points, we need to know the
stabilizer of a given B-type boundary condition. This is rather simple. The 
only candidate fixed point boundary conditions are those with 
$L_i\,{=}\,k_i/2$, for some $i$.  So all we need to do is to determine which
combinations of $f_i$ are allowed phase symmetries. When $n+r$ is odd, it
follows from  \fcurrent\ and \phases\ that every $f_i$ is allowed. Therefore 
the total number  of phase symmetries leaving $\vec L$ fixed is $2^\ell$,
where $\ell$ is the number of $j$ with $L_j\,{=}\,k_j/2$. When $n+r$ is even,
only pairs $f_if_j$ with $i\,{\not=}\,j$ are allowed phase symmetries, 
so the order of the stabilizer is $2^{\ell -1}$. Thus, we see that 
$\nu$ as defined in \nucases\ is indeed precisely given by the order 
of the stabilizer $\cS$ of $\vec L$. 

%%%%%%%%%%%%%%%%%%%%%%%%%%%%%%%%%%%%%%%%%%%%%%%%%%%%%%%%%%%
\section{Projective representation and non-abelian gauge symmetry}
%%%%%%%%%%%%%%%%%%%%%%%%%%%%%%%%%%%%%%%%%%%%%%%%%%%%%%%%%%%

An important result from the general theory \TOAPP\ is that the
number\foot{More precisely, the number of $\ZZ_K$ orbits, where
$K={\rm l.c.m.}(k_i+2)$.} of independent boundary states associated to
a given $\vec L$ is not given by the order $\nu$ of the stabilizer $\cS$,
but rather by the  order $\tilde\nu$ of the {\it untwisted\/} 
(or central) stabilizer $\cU$, which differs from $\nu$ multiplicatively 
by a square number: 
$$
\nu\ =\ \tilde\nu\, N^2 \,.
\eqn\uvss
$$ 
This equation means that a fixed point
boundary state can be resolved into $\tilde\nu$ independent
components that are not further decomposable. It is 
the analogue of the relation $|\Gamma|\,{=}\,\sum_{i=1}^{N_R}
(d_{R_i})^2_{}$ that was derived for orbifolds 
\refs{\DOFI,\fract,\GOMI}, where $\Gamma$ 
is the discrete group that is modded out and $d_{R_i}$ 
is the dimension of the irreducible (projective) representation
$R_i$ of $\Gamma$. In our context the r\^ole of $\Gamma$ is played by the
stabilizer $\cS$, and $\tilde\nu$ is the number of irreducible representations.
In the present BCFT construction, these all have the same dimension 
$N$.\foot{More precisely, when forming tensor products of ordinary
superconformal minimal models, $N=2^{[\tilde\ell/2]}$ 
is always a power of two (see the appendix).  
Other values of $N$, given by powers of $K$, 
should be possible by using $\N\!=\!2$ coset models based on $SU(K)$.}

More concretely, the untwisted stabilizer is associated with an
alternating bihomomorphism (a commutator two-cocycle describing
an element of $H^2(\cS,U(1))$), i.e., a pairing
$$
E:\,\ \cals\times\cals \rightarrow \IC^\times_{}
$$
compatible with the group law and equal to one on the diagonal.
The untwisted stabilizer is
$$
\calu := \{ \Pi\,{\in}\,\cals;\,\, E(\Xi,\Pi)\,{=}\,1\, \forall\;
\Xi\,{\in}\,\cals\}\ .
\eqn\ustab
$$
In general, $E$ is the product of a (not necessarily alternating) 
bihomomorphism $F$ computed from the fixed point modular matrices, 
and the relative monodromies of the currents:\foot{See Appendix A for 
the details of the computation of $F,X,E$ and ${\cal U}$ in Gepner 
models.}
$$
E\ =\ F\,{\rm e}^{2\pi{\rm i} X}\ .
\eqn\torsion
$$
If $E$ is non-trivial 
(which means that ${\nu/\tilde\nu}\!=\!N^2\,{>}\,1$),
we can have only a projective realization 
of the stabilizer group (this may be called ``discrete 
torsion''~\doubref\CVTORS\VW).
The Hilbert space can then be written as $\cH\,{=}\,\cV\otimes\cH'$, 
where $\cV\,{=}\,\IC^N$ is the representation space for an
$N^2$-dimensional projective representation of $\cS$ (the natural
action of $\cS$ on $\cV$ is by multiplication with $N{\times}N$ matrices). 
Accordingly, each of the $\tilde\nu$
vertex operators that describe the emission of $U(1)$ gauge
bosons on the boundary, gains $N{\times}N$ additional
``internal'' indices, thereby forming
vertex operators associated with $U(N)$ gauge symmetry.

The physical picture underlying \uvss\ thus is that the collection of $\nu$ 
gauge fields splits into $\tilde\nu$ separate families, each containing 
$N^2$ gauge fields carrying the adjoint representation of $U(N)$.

Accordingly, the fixed point $D$-brane boundary states split into
$\tilde\nu$ independent ``$N$-fold bound states'', each of which
realizes a $U(N)$ gauge symmetry on its world volume. This is analogous
to the argumentation in  \DOFI\ where $D$-branes on orbifolds with
discrete torsion were considered. Specifically it it was argued
\doubref\DOFI\GOMI\  that discrete torsion in the open string sector
can be attributed to a flat but topologically non-trivial background
$B$-field on a torsion 2-cycle.  The net effect of this is  that the
minimal wrapping number of a $D$-brane is $N$, because configurations
with charge less than $N$  are not allowed \GOMI\ due to global
world-sheet anomalies \doubref\WKTH\FW.  We find that this consistency
condition is naturally encoded in the BCFT, in that the $\tilde\nu$
independent boundary states cannot be decomposed into  further boundary
states with smaller charges.

It would be interesting to more explicitly see how the fixed point
boundary states with $N\!>\!1$ probe the torsion part of the 2-homology
\AMG. More broadly, it may also be possible to give them a meaning in
terms of twisted $K$-theory groups, which seems to be the appropriate
framework for $D$-branes in a $B$-field background \doubref\WKTH\BBACK.
These issues are however beyond the scope of the present paper.

%%%%%%%%%%%%%%%%%%%%%%%%%%%%%%%%%%%%%%%%%%%%%%%%%%%%%%%%%%%
\section{RR charges of resolved boundary states}
%%%%%%%%%%%%%%%%%%%%%%%%%%%%%%%%%%%%%%%%%%%%%%%%%%%%%%%%%%%

We now explain the computation of the RR charges of  the
$B$-type boundary states that arise from resolving fixed points, and
in particular how one finds the corresponding geometric brane charge vectors
$v^{(i)}$, $i\,{=}\,1,...,\tilde \nu$. In general the RR charge is given,
up to a normalization, by the one-point amplitude on the disk with
boundary condition $a$, with the insertion of the bulk vertex operator
$\Phi_{\rm RR}$ of a massless RR state. Since the boundary state
$||a\rangle\rangle$ simply encodes the information
about all such one-point functions, we can write the RR-charge
suggestively as an inner product
$$
q_{\rm RR}(a) \propto \langle \Phi_{\rm RR} \rangle_{a} 
= \langle \Phi_{\rm RR} || a\rangle\rangle
$$
Furthermore, expanding the boundary state in a basis of Ishibashi states,
$$
||a\rangle\rangle = \sum_{i} \cB_{ia} |i\rangle\rangle\ ,
\eqn\ishi
$$ 
we see that the RR charges are  (up to a normalization) nothing else
than the reflection (``Cardy'') coefficients $\cB_{ia}$ of the {\it
massless} RR Ishibashi fields in the expansion of $a$. Of course, the
full expansion contains many more terms, but those correspond to
massive, non-topological components that we are not interested in.

The problem with this computation is that the resulting charges are not
properly normalized and that there is no immediate connection to a
geometric basis of the  charge lattice. Both normalization and basis
can be fixed by  computing the intersection form, given in CFT by the
index $tr(-1)^F$, and comparing it with the geometric intersection form
at the Gepner point \BDLR. It was found in \BDLR\ that there is a very
simple relation  between the intersection numbers of the $L_i=0$
states and the  intersection form of the periods in the
$\ZZ_K$-symmetric basis. Thus, once the ambiguities  are fixed for the
$\vec L=0$ states,  the charges of all remaining boundary states, and
in particular of the resolved ones, can be determined from the 
reflection coefficients.

For the reflection coefficients of the resolved $B$-type boundary 
states (constructed as $A$-type states in the mirror modular invariant
associated to the simple current group, as explained above) we
find explicitly: 
$$
\eqalign{
\cB_{(\lambda,\mu,f)(\vec L,M,S,\Psi)} &=
\sqrt{\frac{ |{\cal G}^{\rm mirr}|}{{\nu}_{\vec L} {\tilde\nu}_{\vec L}}}\,
     \Psi(f)
\prod_{i\notin I_f} 
2\sqrt{\frac{2}{h_i}}\, \sin\!\left[\pi\frac{(l_i+1)(L_i+1)}{h_i}\right]
\cr &\;\;\times 
\prod_{i\in I_f}
\ee^{-2\pi\ii\, 3k_i/16}  
%\cr
\prod_{i=1}^r
\frac{1}{\sqrt{2h_i}}\, \ee^{2\pi\ii M m_i t_i/2h_i} \;\;
\frac{1}{2^r}\, \ee^{-2\pi\ii (S s_1 + \sum_{i=2}^r S^2 s_i ) /4}
}\eqn\cardy
$$
In this expression, $(\lambda,\mu,f)$ labels the Ishibashi states,
which according to the general theory requires  $f\,{=}\,\prod_{i\in I_f}
f_i$ to be a simple current which fixes the bulk field label
$(\lambda,\mu)\,{=}\,(l_1,...,l_r,m_1,...,m_r,s_1,...,s_r)$. (In most cases,
$f$ will simply be the identity field, but non-trivial $f$'s are
possible when $l_i\,{=}\,k_i/2$ for some $i$). The combination
$(\lambda,\mu,f)$ must  in addition possess the right relative
monodromy to cancel the discrete  torsion with respect to all
symmetries of the Gepner-Greene-Plesser construction outlined above.
Furthermore, $\Psi$ is a character of the untwisted stabilizer  of
$\vec L$, $M\,{=}\,0,...,2K{-}1$ measures the unbroken spacetime
supersymmetry\foot{$M$ runs over even or odd values depending 
on the parity of $\vec L$.}, and $t\,{=}\,(t_1,...,t_r)$ is the 
combination with minimal (non-integer) $U(1)$ charge, 
i.e.\ $\sum_i t_i/h_i\,{=}\,1/K$. The label $S$ distinguishes 
branes ($S\,{=}\,0,1$) from anti-branes ($S=2,3$). Moreover, a
crucial ingredient is the factor $\ee^{-2\pi\ii 3k_i/16}$ which comes
from the resolution of the fixed points. Its form is inferred from
known fixed point matrices \refs{\BSSY,\FUSSF} for the simple 
currents in su$(2)$ WZW models. Finally, $|{\cal G}^{\rm mirr}|$ is
the order of the Gepner-Greene-Plesser simple current group.

Note that formula \cardy\ gives the minimal $D$-brane charges, namely
the charges when there is no discrete torsion. As mentioned above, when
there is discrete torsion, there are ``$N$-fold bound states'' which
cannot be decomposed further, and consistency \FW\ requires that the
allowed charges are an integral multiple of \cardy, i.e.,
$$
Q\ =\ N\,Q_{\rm min}\ =\ N\,\cB\ .
\eqn\Qnorm
$$
This is the BCFT analog of the formula $Q=d_R/|\Gamma|$ for
$D$-branes on orbifolds with discrete torsion \fract.

%%%%%%%%%%%%%%%%%%%%%%%%%%%%%%%%%%%%%%%%%%%%%
\chapter{An example}
%%%%%%%%%%%%%%%%%%%%%%%%%%%%%%%%%%%%%%%%%%%%%

We consider the Gepner model with $(k_1,k_2,k_3)\,{=}\,(4,4,4)$,
which geometrically corresponds to a $K3$ defined by the equation
$\sum_{i=1}^3 {x_i}^6+{x_4}^2\,{=}\,0$ in $\IP(1,1,1,3)[6]$. 
It figures as fiber in a CY threefold that was investigated in 
\doubref\KLLW\ES. In these references, the appearance of 
brane configurations with $\nu\,{=}\,1,2,4,8$ was noticed,
and this was one of the motivations for the present investigation.

In order to determine the RR charge vectors from the reflection coefficients
\cardy, we first need to have the complete set of massless Ishibashi
RR states. For the model at hand, the following massless RR fields can 
couple to B-type boundary states:
$$
\eqalign{
\rr1\ &=\ \Big[\phi^L_{(0,1,1)} \phi^R_{(0,-1,-1)}\Big]^3 \cr
\rr2\ &=\ \Big[\phi^L_{(0,-1,-1)} \phi^R_{(0,1,1)}\Big]^3 \cr
\rr3\ &=\ \Big[\phi^L_{(2,3,1)} \phi^R_{(2,-3,-1)}\Big]^3 \cr
}\eqn\Rgs
$$
where $\phi^{L,R}_{(l,m,s)}$ are left- and right-moving primary fields.
Using \cardy, we find that the charges of the orbit of 
$\vec L\,{=}\,(0,0,0),\;S\,{=}\,0$ states in this basis are, up to 
normalization ($\rho := \ee^{2\pi\ii /6}$):
%$$
%\pmatrix{ 1 & 1 & 1 \cr
%\rho & \rho^5 & \rho^3 \cr
%\rho^2 & \rho^4 & \rho^6 \cr
%\rho^3 & \rho^3 & \rho^3 \cr
%\rho^4 & \rho^2 & \rho^6 \cr
%\rho^5 & \rho & \rho^3 }
%
$$
Q^{\rm BCFT}_{(0,0,0)} =
\pmatrix{1 & 1 & 1 \cr
\rho & -\rho^2 & -1 \cr
\rho^2 & -\rho & 1 }\ .
\eqn{\Qooo}
$$
The columns correspond to the three Ramond ground states \Rgs,
while the rows correspond to the different $M$ labels, i.e.\
$M\,{=}\,0,2,4$ for $\vec L\,{=}\,(0,0,0)$. We have suppressed the lower half 
of this matrix ($M\,{=}\,6,8,10$), because it is simply the negative of 
the upper half and so represents the anti-branes. From \KLLW, we know the
analytic continuation from the Gepner to the geometric charge basis
at large radius, and by inverting \Qooo, we can easily find the
matrix $H$ that furnishes the change of basis:
$$
H = \frac13
%\pmatrix{ \rho^2 & -\rho^2-\rho & 2 \cr
%-\rho& \rho^2+\rho & 2 \cr
%4 & 0 & -4 }
\pmatrix{ \rho^2 & 2 & -\rho \cr
-\rho&  2 & \rho^2\cr
4 & -4 & 4 }\ .
\eqn\Hma
$$
Indeed, multiplying \Qooo\ from the right with $H$, we obtain,
$$
v_{(0,0,0)}^i\ =\ Q^{\rm BCFT}_{(0,0,0)} H\ =\ 
\pmatrix{ 1 & 0 & 1\cr
-2 & 2 & -1 \cr
1 & -2 & 2 } \,,
$$
which gives back the charges of the $L_i\,{=}\,0$ states
as given in \doubref\KLLW\ES\ (up to a slight change of basis).

To compute the charges of the remaining states, we notice
that the ambiguity in the normalization of \Qooo\ rests inside each 
column. Also, compared to equation \cardy, we have omitted the
factors $\bigl( \sin^3\frac{\pi}{6},\sin^3\frac{\pi}{6}, 
\sin^3\frac{3\pi}{6}\bigr)$.
Such ambiguities just change the normalization of each charge, 
and can be adjusted by a redefinition of the basis change 
\Hma\ that connects the BCFT charges with the geometric brane 
charges. With this normalization in mind, one can easily 
compute the charges of the boundary states with 
${\vec L}\,{\not=}\,(0,0,0)$. With the help of trigonometric 
identities one can thereby recover the charges 
found in \doubref\BDLR\DDCR, and listed in the Table below.

For example, the charges of states with 
$\vec L=(2,2,2)$ are given by
$$
\eqalign{
Q_{(2,2,2)}^{\rm BCFT} &=
%\frac12
\pmatrix{
\left(\frac{\sin{\frac{3\pi}{6}}}{\sin{\frac{\pi}{6}}}\right)^3 &
\left(\frac{\sin{\frac{3\pi}{6}}}{\sin{\frac{\pi}{6}}}\right)^3 &
\left(\frac{\sin{\frac{9\pi}{6}}}{\sin{\frac{3\pi}{6}}}\right)^3 \cr
\left(\frac{\sin{\frac{3\pi}{6}}}{\sin{\frac{\pi}{6}}}\right)^3 \rho  &
\left(\frac{\sin{\frac{3\pi}{6}}}{\sin{\frac{\pi}{6}}}\right)^3 \rho^5 &
\left(\frac{\sin{\frac{9\pi}{6}}}{\sin{\frac{3\pi}{6}}}\right)^3 \rho^3 \cr
\left(\frac{\sin{\frac{3\pi}{6}}}{\sin{\frac{\pi}{6}}}\right)^3 \rho^2  &
\left(\frac{\sin{\frac{3\pi}{6}}}{\sin{\frac{\pi}{6}}}\right)^3 \rho^4 &
\left(\frac{\sin{\frac{9\pi}{6}}}{\sin{\frac{3\pi}{6}}} \right)^3 
} \cr
&= 
\pmatrix{
8 & 8 & -1 \cr
8\rho & -8\rho^2 & 1 \cr
8 \rho^2 & -8\rho & -1
}
=
(g^{-1}+1+g)^3\, Q_{(0,0,0)}^{\rm BCFT}\ ,
}
\eqn\Qttt
$$
where $g=\pmatrix{0&1&0 \cr 0&0&1 \cr -1&0&0}$ is the appropriate shift
matrix. This yields the following RR charges in the geometrical basis:
$$
v_{(2,2,2)}^i\ =\ Q^{\rm BCFT}_{(2,2,2)} H\ =\ 
\pmatrix{
-4 & 12 & -4\cr
-4  & 4 & 4 \cr
-4 & -4 & 4
}\ .
\eqn\chgeo
$$
More precisely, because of the fixed points, these charges are the
unresolved {\it overall} charges of reducible brane configurations. We
now would like to know into  how many irreducible components these
(and also the other $\nu>1$) states split,  
and what the charge vectors $v$ of these components are. In
the following, we will explicitly work out the relevant combinatorics
of the twisted and untwisted stabilizers in our example. More details
regarding the general case can be found in the appendix.

The phase symmetries \phases\ we divide by in the Greene-Plesser 
construction are generated by the following simple currents:
$$
\eqalign{\Pi_1 &= p_1 p_2^{5} \cr
\Pi_2 &= v_1 p_1^3\ .}
\eqn\phassymm
$$
As is easy to see, and follows from the general discussion after 
eq.\ \phases, the simple current group generated by the currents
\phassymm\ together with the $w_i$ and $u=v_1p_1p_2p_3$, contains
all three currents $f_1,f_2$, and $f_3$. Because of the permutation
symmetry, there are then three different possible
stabilizers in our example:
$$
\cals=\cases{
\ZZ_2=\{1,f_1\} & $\vec L = (2,*,*)$ $(\nu=2)$\cr
\ZZ_2^2=\{1,f_1,f_2,f_1f_2\} & $\vec L= (2,2,*)$ $(\nu=4)$\cr
\ZZ_2^3=\{1,f_1,f_2,f_3,f_1f_2,f_1f_3,f_2f_3,f_1f_2f_3\} & $\vec L=(2,2,2)$ 
($\nu=8)$}
\eqn\stabs
$$
(where $*=0,1$).

To determine the untwisted stabilizer, we need to determine 
the bihomomorphism \torsion. 
In the case of our interest, all levels are equal to 
zero modulo $4$. It is then easy to see that the fixed point matrices 
in \cardy\ satisfy the same simple current relations as the modular
S matrix \BSSY. Therefore, the bihomomorphism $F$ which would
measure the deviation from the usual simple current relations,
is identically equal to one. Furthermore, the matrix of relative
monodromies on the stabilizer is given by
$$
X(f_i,f_j) = X(p_i^{h_i/2},p_j^{h_j/2}) + X(v_i,v_j)
= -\frac{\delta_{ij}}{h_i} \frac{h_i}2\frac{h_j}2 + \frac12 =
\cases{
\ 0 \bmod\ZZ & $i=j$ \cr
1/2 \bmod\ZZ& $i\not= j$}
$$
on the generators of the stabilizers. Thus, $E$ is given by
$$
E(f_i,f_j) = (-1)^{1+\delta_{ij}} \,.
$$
Consider the untwisted stabilizer \ustab\ for the case $\nu\,{=}\,4$
in \stabs. Since $E(f_1,f_2)\,{=}\,E(f_2,f_1)=E(f_1f_2,f_1)
\,{=}\,{-}1$, we
see that in fact no non-trivial element of the stabilizer $\ZZ_2^2$ 
is in the kernel of $E$, and the untwisted stabilizer is trivial. More
generally we find for all the other cases:
$$
\calu = \cases{
\ZZ_2=\{1,f_1\} \ ,\ \ \qquad\tilde\nu=2 & (for $\nu=2$) \cr
\{{\rm id} \} \ ,\ \qquad\qquad\qquad \tilde\nu=1 & (for $\nu=4$) \cr
\ZZ_2=\{1,f_1f_2f_3\} \ ,\ \ \ \tilde\nu=2 & (for $\nu=8$)$\;$,}
$$
which corresponds to $N\,{=}\,1,2,2$, respectively, in \uvss.
According to our general reasoning, we thus find non-abelian
gauge groups for the $\nu\,{=}\,4,8$ boundary states, namely
$U(2)$ and $U(2)\times U(2)$, respectively.

The general result, computed in the appendix, is
that for Gepner models, $\tilde\nu\,{=}\,1$ when $\tilde\ell$ is 
even, $\tilde\nu\,{=}\,2$ when $\tilde\ell$ is odd, and hence
$N\,{=}\,2^{[{\tilde l}/{2}]}$.

To determine the resolved charge vectors from equation \cardy, 
we need to know the complete labels of the 
RR ground states \Rgs, i.e., not
only the bulk labels, but also the currents associated with them.
Now $\rr1$ and $\rr2$ obviously have trivial stabilizer, so 
they must be combined with the identity simple current. However,
for $\rr3$ the stabilizer is $\ZZ_2\,{=}\,\{1,f_1f_2f_3\}$, and
there is a slight ambiguity as to what we mean actually by
$\rr3$. By considering the fibration of the K3 to a CY threefold,
it appears that the only consistent choice is to associate
$\rr3$ with the identity as well.

Having fixed the labels, and taking into account \cardy\ and 
\Qnorm, we then find that the charge vectors of the resolved
boundary states 
are simply given by $1/\tilde\nu$ times the charge vectors
of the unresolved states. That is, the coefficients 
of their expansion into massless RR Ishibashi states \ishi\ 
turn out to be the same (up to an overall factor of $1/\tilde\nu$) as
for the unresolved boundary states.\foot{Due to the
independence of the $\tilde\nu$ resolved states, these must
then differ in the massive, non-topological Ishibashi expansion components.}
This is fortunate, since, as we will discuss in the next section,
it is precisely what we expect from the geometry of semi-stable sheaves.

Note that the untwisted stabilizer (i.e., the discrete torsion) plays
an important r\^ole in determining what the charges of the resolved
$D$-brane states are. For example, for the fixed point boundary
states with $\nu=8$ the
unresolved charges $v_{(2,2,2)}^i$ in \chgeo\ are multiples of four,
and one might have been tempted to believe that the resolved states
have $1/4$ of these charges. However, the resulting charges cannot
describe physical states, as the complex dimension of the moduli
spaces, given by the Mukai formula $\mu(v)\,{=}\,\langle v,v\rangle+2$
\MUKAI, turns out to be fractional; in other words, the charges are not
properly quantized. This is a reflection of the fact, as mentioned
above, that in the presence of discrete torsion,
${\nu/\tilde\nu}\!\equiv\!N^2\,{>}\,1$, we have ``$N$-fold bound
states'' whose  charges are $N$ times larger (cf., \Qnorm); 
for the $\nu=8$ states we have $\tilde\nu=2$ so that 
the resolved states have charges which are 
one-half of $v_{(2,2,2)}^i$ in \chgeo.

We have summarized all 
the relevant data in Table~2, which is the refinement of a similar 
table in ref.\ \KLLW, where the subtleties of the untwisted stabilizer
were not taken care of.

\goodbreak
\def\ft[#1,#2,#3]{(#1,#3,#2)}
\def\fl[#1,#2,#3]{\ \ [#1,#2,#3]\ \ }

\vskip 1.truecm
\begintable
$L_i$  |\qquad & $\!\!\!\!\!\!\!\!\!\!\!\!\!\!\!\!\!\!\!\!v(\cE)=
\big(r,c_1(\cE),r{+}{1\over2}{c_1}^2(\cE){-}c_2(\cE)\big)
\!\!\!\!\!\!\!\!\!\!\!\!\!\!\!\!\!\!\!\!$ &\qquad | \ $\nu$ \
& \ $\tilde\nu$ \ | \ $G$\ \crthick
\fl[0,0,0]  | \ft[1,1,0]  & \ft[1,2,-2]  & \ft[2,1,-2] | 1 & 1 |
\ $U(1)$\ \cr 
\fl[1,0,0]  | \ft[1,-1,0] & \ft[1,0,-2] & \ft[0,-1,2]  | 1 & 1 |
\ $U(1)$\ \cr 
\fl[1,1,0]  | \ft[1,1,-4] & \ft[1,-2,2]  & \ft[2,-1,-2] | 1 & 1 |
\ $U(1)$\ \cr
\fl[1,1,1]  | \ft[3,-3,0] & \ft[3,0,-6] & \ft[0,-3,6]  | 1 & 1 |
\ $U(1)$\ \crthick 
\fl[2,0,0]  | \ft[2,0,0] & \ft[2,2,-4]  & \ft[0,-2,0]  | 2 & 2 |
\ $U(1)\times U(1)$\ \cr
\fl[2,1,0]  | \ft[2,-2,0] & \ft[2,0,-4] & \ft[0,-2,4]  | 2 & 2 |
\ $U(1)\times U(1)$\ \cr 
\fl[2,1,1]  | \ft[2,2,-8] & \ft[2,-4,4]  & \ft[4,-2,-4] | 2 & 2 |
\ $U(1)\times U(1)$\ \crthick  
\fl[2,2,0]  | \ft[4,0,-4] & \ft[0,-4,4]  & \ft[0,0,4]  | 4 & 1 |
\ $U(2)$\ \cr
\fl[2,2,1]  | \ft[4,-4,0]  & \ft[4,0,-8] & \ft[0,-4,8] | 4 & 1 |
\ $U(2)$\ \crthick 
\fl[2,2,2]  | \ft[4,4,-12] & \ft[4,-4,-4] & \ft[4,-4,4] | 8 & 2 |
\ $U(2)\times U(2)$\ 
\endtable
%\vskip-10pt
\noindent{\bf Table 2:}
{\sl
Labels and unresolved RR brane charges of boundary states on
the $K3$ surface in $\IP(1,1,1,3)[6]$. 
Furthermore,  $\nu$ denotes the order of the stabilizer of
the fixed points, which gives the total number of gauge fields, while
$\tilde\nu$ is the order of the untwisted stabilizer, which gives the
number of $U(N)$ factors and 
irreducible components (with charges given by $1/\tilde\nu$
of the overall charges). On the right we list the unbroken gauge
groups $G$, as implied by the discrete torsion. Geometrically, 
configurations with $\tilde\nu\,{>}\,1$ correspond to strictly 
semi-stable sheaves.
} 

%%%%%%%%%%%%%%%%%%%%%%%%%%%%%%%%%%%%%%%%%%%%%
\chapter{Boundary fixed points and semi-stable sheaves} 
%%%%%%%%%%%%%%%%%%%%%%%%%%%%%%%%%%%%%%%%%%%%%

We now discuss the resolution of the boundary CFT fixed points from the
viewpoint  of space-time geometry. Note first of all that the simple
currents $f_i$ generically have non-integer dimensions, and when this
is the case, the fixed points cannot appear in the bulk, but only on
the boundary. Geometrically this should mean that these fixed points
do not correspond to singularities of the manifold, but
rather to the degeneration of bundles over it.

This touches upon an interesting mathematical issue, namely the
compactification of the moduli space of instantons  (mathematically:
holomorphic vector bundles, or more generally torsion-free coherent
sheaves)  on $K3$. Recall that for a given Mukai charge vector $v(\cE)$
\Mukdef, the complex dimension of the moduli space is  $\mu(v(\cE))=
\langle v(\cE),v(\cE)\rangle+2$ \MUKAI. Over this space the structure
of bundles or sheaves $\cE$ changes, and can in particular degenerate.

To be specific, consider for example rank two configurations with
charges\foot{The compactification of the moduli space of such bundles has
been investigated in ref.~\OGRADY.} $v\,{=}\,(2,0,2{-}2k)$, $k\,{\geq}\,2$
(which indeed appear in the BCFT construction of the $K3$ we discussed
in the present paper, see Table 2 and \doubref\KLLW\ES). These
correspond to $SU(2)$ bundles with instanton number $c_2\,{=}\,2k$, whose
moduli spaces have dimension $\mu\,{=}\,2rc_2-2\,{\rm dim}\,G\,{=}\,8k-6$.
As is well-known, these moduli spaces are (almost) the same as the moduli
spaces of $SU(2)$ $\N\,{=}\,2$ gauge theories with $N_m\,{=}\,2rc_2$
hypermultiplets. At generic points the Higgs VEVs break the gauge
symmetry completely ($\nu\,{=}\,\tilde\nu\,{=}\,1$); however at the 
origin, where the VEV's of all hypermultiplets vanish, there is an extra 
unbroken $SU(2)$ gauge group ($\nu\,{=}\,4,\;\tilde\nu\,{=}\,1$). In terms
of the brane picture, this
degeneration corresponds to small instantons, ie., point-like
$D0$-branes. One can also envision a degeneration where just an extra
$U(1)$ factor appears, which then would correspond to a reducible
configuration with $\nu\,{=}\,\tilde\nu\,{=}\,2$. In particular, the 
degeneration into a reducible line bundle, $\cE\sim L\oplus L^{-1}$, 
was discussed in \BS. Physically, this means that the brane configuration 
splits into two irreducible components with non-zero first Chern class:
$v^{(\pm)}\,{=}\,(1,\pm c_1,1{-}k)$.

In the present paper, we have found that resolving  BCFT fixed points
amounts to decomposing a reducible brane system into irreducible 
components in the simplest manner, 
namely into building blocks with identical charges.
Specifically, what we find is, for example, that a rank two bundle with
charges $(2,0,2-2k)$  and $\nu\,{=}\,2$ splits into two configurations, 
each with charges $(1,0,1-k)$. Unlike the above-mentioned degeneration
$v\to v^{(+)}\oplus v^{(-)}$ \BS, such a degeneration cannot be
described in terms of ordinary $U(1)$ line bundles (since these would
imply $c_1\,{\not=}\,0$), but it does not have to.

Rather, it is known \GIES\ that the moduli space of stable bundles on
$K3$ is naturally compactified by adding strictly semi-stable sheaves,
which is more general than bundles.\foot{For related considerations in
the physics literature, see e.g., \refs{\HM,\CV,\DM,\DFRa,\BFMM}.}
Physically, this amounts to including point-like degrees of freedom, in
the present example ($k-1$) $D0$-branes  sitting on each of two
$D4$-branes.\foot{Since this configuration is expected to exist also
when the $K3$ is large, this should have a conventional interpretation,
for example in terms of a background $B$-field.}
Such strictly (Gieseker) semi-stable sheaves $\cE$ have the 
property \MUKAI\ that they have proper sub-sheaves $\cE'$ 
with ${\rm rk}\,\cE'\,{<}\,{\rm rk}\,\cE$ and charges $v(\cE')$, 
such that the normalized Mukai vectors $v(\cE)/{\rm rk}\,\cE$
and $v(\cE')/{\rm rk}\,\cE'$ are equal. This can only occur
%\foot{Non-coprime charges do not imply that a brane
%configuration must be semi-stable. Rather, there are then singular loci
%somewhere in the moduli space over which the brane configuration
%degenerates, which in CFT language can be recognized by $\nu>1$. For
%example, we have found (see Table~2)
%a semi-stable configuration with $v=(3,0,-3)$ and
%$\nu=1$, which represents a perfectly smooth, irreducible $SU(3)$
%bundle.} 
when \GCD$(q_4,q_2,q_0)\,{>}\,1$ (for primitive charge vectors,
\GCD$(q_4,q_2,q_0)\,{=}\,1$, the moduli space is already compact).
Physically this condition corresponds to collinear central
charges and thus to bound states at threshold. 

Precisely this structure, namely the degeneration into components with
charges $v^{(i)}\,{=}\,v(\cE)/\tilde\nu$, is what we find from
CFT. While expected on general grounds, this is nevertheless
reassuring, since a priori the resolution of boundary fixed points
might have given something else and turned out to be incompatible with
a geometric picture, in particular with the picture of decomposing into
sub-sheaves. We thus meet another instance where abstract properties
of 2d superconformal field theory possess an identifiable, concrete 
geometrical meaning when translated into the space-time picture.

%%%%%%%%%%%%%%%%%%%%%%%%%%%%%%%%%%%%%%%%%%%%%%%%%%%%%%%
\ack
%%%%%%%%%%%%%%%%%%%%%%%%%%%%%%%%%%%%%%%%%%%%%%%%%%%%%%%
\nobreak
We are grateful to L.R.\ Huiszoon, C.\ R\"omelsberger 
and A.N.\ Schellekens for useful discussions.
J.W. would like to thank J\"urg Fr\"ohlich for helpful
discussions. J.F., W.L, C.S.\ and J.W.\ would like to
thank ESI (Vienna) for hospitality.

%%%%%%%%%%%%%%%%%%%%%%%%%%%%%%%%%%%%%%%%%%%%%
\appendix{A}{Untwisted stabilizers}
%%%%%%%%%%%%%%%%%%%%%%%%%%%%%%%%%%%%%%%%%%%%%

We discuss the general fixed point combinatorics for B-type 
boundary states in Gepner models and in particular derive the 
general rule $\tilde\nu\,{=}\,2^{\frac{1-(-1)^{\tilde\ell}}{2}}$ 
for the order of the untwisted stabilizer. Together with 
$\nu\,{=}\,2^{\tilde\ell}$, this implies $N\,{=}\,2^{[{\tilde\ell}/{2}]}$.

We first recall from \FUSSF\ the definition of the bihomomorphism 
$F$ on the stabilizer. Consider some CFT with modular S matrix 
$S_{ab}$ and some simple current group ${\cal G}$. For all $J\,{\in}
\,{\cal G}$, one can define a fixed point matrix $S^J_{ab}$ between 
fields $a,b$ with $J\,{\in}\,{\cal S}_a\,{\cap}\,{\cal S}_b$, where 
${\cal S}_a\,{\subseteq}\, {\cal G}$ 
is the stabilizer of the primary field $a$. While the S matrix
satisfies the usual simple current relation
$$
S_{a,Kb} = \ee^{2\pi\ii Q_K(a)} S_{ab}\;,
$$
where $Q_K(a)$ is the monodromy charge of $a$ with respect to $K$,
this is generically violated for the fixed point matrices, and
the violation is measured by $F$, i.e.:
$$
S^J_{a,Kb} = \ee^{2\pi\ii Q_K(a)} S_{ab}^{}\, F^*_{b}(K,J)\ .
$$

The second ingredient in the bihomomorphism \torsion\ is
the pairing $X$ (defined modulo $\ZZ$). The symmetric part of
$X$ is determined by the conformal weights of the currents. 
According to the general results of \KRSC, the antisymmetric 
part of $X$ can be freely chosen (with some restrictions) 
and together with the choice of simple current group ${\cal G}$ 
determines the modular invariant\foot{Notice that in \KRSC, 
the antisymmetric part of $X$ is called ``discrete torsion''
on the simple current group. In the main part of the present 
paper, we have chosen to call $E$ the ``discrete torsion'' on 
the stabilizer.}.

We now compute $F$ in the situation of the main text, 
considering the case $n+r$ even first.
For a given $\vec L$, we denote $I\,{:=}\,\{i;L_i{=}k_i/2\}$,
distinguish some $a_0\,{\in}\,I$ (assuming $I\,{\not=}\,\emptyset$), 
denote the corresponding simple
current by $f_0$, and let the stabilizer $\cals_{\vec L}$ be
generated by $f_{0a}\,{=}\,f_0f_a$, with $a\,{\in}\,I$.
We have $\cals_{\vec L}\,{\cong}\,(\zet_2)^{\ell-1}$, where
$\ell\,{=}\,|I|$. To compute the twisting of simple current relations,
consider $(\vec L,M,S)\,{=}\,(..,k_0/2,..,k_a/2,..,k_b/2,..,...)$ and
$(\lambda,\mu)\,{=}\,(...,k_0/2,..,k_a/2,..,l_b,...,...).\,$%
\foot{We only divide $k$'s by two when they are even.} 
Explicitly, the relevant $SU(2)$ part of the fixed point matrix 
is, up to irrelevant factors,
$$
S^{f_{0a}}_{\lambda,\vec L}
\sim \prod_{i\not= a_0,a}
\sin\!\left[\pi\frac{(l_i+1)(L_i+1)}{h_i}\right]
 \prod_{i=a_0,a} \ee^{-2\pi\ii\, 3k_i/16} \;,
$$
which is a part of the expression \cardy. Then one easily finds:
$$
S^{f_{0a}}_{\lambda, f_{0b}\vec L} =
\cases{
(-1)^{l_b} S^{f_{0a}}_{\lambda,\vec L} & $a\not= b$ \cr
S^{f_{0a}}_{\lambda,\vec L} & $a=b
\;.$}
$$
We conclude that
$$
F_{\vec L}(f_{0b},f_{0a}) = 
\cases{
(-1)^{k_0/2} & $a\neq b$ \cr
(-1)^{k_0/2+k_a/2} & $a=b
\;$,}
$$
which gives the first part of \torsion.

To find the correct choice of $X$ requires a careful
analysis of the Greene-Plesser construction. In turns out
that to obtain the mirror modular invariant as a simple 
current invariant, one has to define $X$ by 
$X(p_a,p_b)\,{=}\,\delta_{ab}h_a/2$ and $X(v_a,v_b)\,{=}\,1/2$.
Recalling $f_a\,{=}\,p_a^{h_/2} v_a$, one finds:
$$
X(f_{0a},f_{0b}) = \frac{h_0}{4} + \frac{h_a}{4}\, \delta_{ab}\;.
$$
Putting things together, we thus have:
$$
E_{\vec L}(f_{0a},f_{0b}) = F_{\vec L}(f_{0a},f_{0b})\,
{\ee}^{2\pi \ii X(f_{0a},f_{0b})} =
(-1)^{1+\delta_{ab}}\ .
$$

Computing the untwisted stabilizer \ustab\ is now an easy exercise.
Consider some $\phi\,{=}\,\prod_{b\in I'} f_{0b}\,{\in}\,{\cal S}_{\vec L}$,
with $I'\,{\subseteq}\,I\,{\setminus}\{a_0\}$.
$$
E_{\vec L}(f_{0a},\phi) = 
\cases{
(-1)^{|I'|} & $a\,{\notin}\, I'$ \cr
(-1)^{|I'|-1} & $a\,{\in}\, I'
\;$.}
$$
For $\phi$ to be in $\calu_{\vec L}$, we must require 
$E_{\vec L}(f_{0a},\phi)\,{=}\,1$ for all $a$. This is only possible 
if $I'\,{=}\,\emptyset$, or if $I'\,{=}\,I\,{\setminus}\{a_0\}$ and 
$|I|\,{=}\,\ell$ is even. We conclude:
$$
\calu_{\vec L} = 
\cases{
\zet_2 & $\ell$ even, $\ell\not= 0$ \cr
\{{\rm id}\} & $\ell$ odd or $\ell = 0
\;$.}
$$

The combinatorics for the boundary states in the case 
$n\,{+}\,r\,{=}\,$odd can be mapped to $n\,{+}\,r\,{=}\,$even by 
appending a trivial factor with $k_0\,{=}\,0$. Put differently, 
the above derivation still holds by letting $\cals_{\vec L}$ be 
generated by $f_{0a}\,{=}\,f_a$, without distinguishing any 
particular $a\,{\in}\,I$. Simple current twists and monodromy carry 
over mutatis mutandis, but the final result is somewhat different:
$\phi\,{\in}\,\calu_{\vec L}$ if either $I'\,{=}\,\emptyset$ or 
$I'\,{=}\,I$, and $|I|$ odd. We conclude:
$$
\calu_{\vec L} = 
\cases{
\zet_2 & $\ell$ odd \cr
\{{\rm id}\} & $\ell$ even 
$\;$.}
$$
The claims then follow.

\bigskip
\goodbreak
\refout
\end